\documentclass[twocolumn,showpacs,prl]{revtex4}
\usepackage{bm}
\usepackage{epsfig}
\newcommand{\nix}[1]{}

\begin{document}

\title{Symmetry and spin dephasing in (110)-grown quantum wells}
\author{V.V.~Bel'kov$^{1,2}$, P.~Olbrich$^1$, S.A.~Tarasenko$^2$, 
D.~Schuh$^1$, W.~Wegscheider$^1$, T.~Korn$^{1}$,   Ch.~Sch{\"u}ller$^{1}$, 
D.~Weiss$^{1}$, W.~Prettl$^1$, and S.D.~Ganichev$^{1}$
}
\affiliation{$^1$Terahertz Center, University of Regensburg,
93040 Regensburg, Germany}
\affiliation{$^2$A.F.~Ioffe Physico-Technical Institute, Russian
Academy of Sciences, 194021 St.~Petersburg, Russia}

\begin{abstract}
Symmetry and spin dephasing of in (110)-grown GaAs quantum wells (QWs) are investigated applying 
magnetic field induced photogalvanic effect (MPGE) and time-resolved Kerr rotation.
We show that MPGE 
provides a tool to probe the  symmetry of (110)-grown quantum wells. 
The photocurrent is only observed for asymmetric structures 
but vanishes for symmetric QWs. Applying  Kerr rotation we prove
that in the latter case the spin relaxation time is maximal,
therefore these structures set upper limit of
spin dephasing in GaAs QWs. We also demonstrate that structure 
inversion asymmetry can be controllably tuned 
to zero by variation of delta-doping layer position.
\end{abstract}

\maketitle

Quantum wells  on (110)-oriented GaAs substrates attract 
growing attention in spintronics due to their extraordinary slow spin dephasing~\cite{Ohno1999,Harley03,x3,x4}. 
The reason for the long spin lifetime of several nanoseconds is the (110) crystal orientation: 
Then the effective magnetic field due to spin-orbit coupling points into 
the growth direction~\cite{DK86} and spins oriented along this direction do not precess. 
Hence the D'yakonov-Perel' spin relaxation mechanism~\cite{DP71} which is based 
on the spin precession in the effective magnetic field and
usually limits the spin lifetime of conduction electrons is suppressed.
If however QWs are asymmetric the structure inversion symmetry is broken and 
Rashba spin-orbit coupling causes an in-plane effective magnetic field, 
thus speeding-up spin dephasing. To judge the symmetry of QWs one has to rely on 
the  growth process but no independent method 
to check the structure symmetry is readily available. 
Here we show that the magneto-photogalvanic effect~\cite{Belkov05} 
is an ideal tool to probe the  symmetry of (110)-grown QWs. 
The photocurrent is only observed for asymmetric structures but vanishes 
if QWs are symmetric. We further show via time-resolved Kerr rotation 
that the spin relaxation time is maximal whenever the QW is symmetric.

\begin{figure}[t]
\centerline{\epsfxsize 70mm \epsfbox{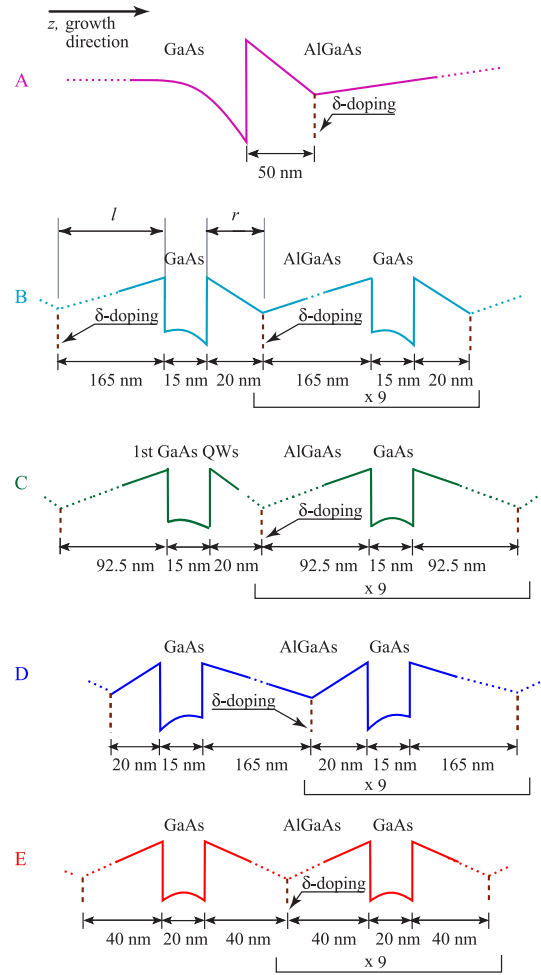}}
\caption{ Band profile of QWs and doping position.}
\label{fig1}
\end{figure}

Absorption  of light in a quantum well in the presence 
of a magnetic field can generates a photocurrent. This is called as 
magneto-photogalvanic effect~\cite{Ganichev2006,Ivchenkobook2}. 
The microscopic model of the MPGE in QWs having bulk inversion 
asymmetry (BIA) and/or structure inversion 
asymmetry (SIA) is based on the asymmetry of photoexcitation 
and relaxation~\cite{naturephysics06}. 
So far it has been observed in (001)-grown GaAs, SiGe and  InAs QWs~\cite{Belkov05,naturephysics06,SiGe07}. 
The  experiments here are carried out on MBE (110)-grown 
Si-$\delta$-doped n-type  GaAs$/$Al$_{0.3}$Ga$_{0.7}$As  structures 
(see Table~\ref{table1}).
The conduction band profile of the investigated structures 
together with the corresponding $\delta$-doping position is shown in Fig.~\ref{fig1}. QWs
differ essentially  in their doping profile: 
Sample A is a single heterojunction and has the strongest asymmetry 
stemming from the triangular confinement potential. In samples B and D, the doping layers are
asymmetrically shifted off the barrier center either to the left  or to
the right, respectively (see Fig.~\ref{fig1}). This asymmetric doping
yields an asymmetric potential profile inside the QWs. To describe the degree of asymmetry 
we introduce the parameter $\chi = (l-r) / (l+r)$, where
$l$ and $r$ is the  spacer 
layer thickness between QW and $\delta$-doping position 
(see Fig.~\ref{fig1} and Table~\ref{table1}).
Samples C and E contain a Si-$\delta$-sheet, placed in the center of each barrier  
between adjacent  QWs. While  sample E 
was grown fully symmetrical, a weak asymmetry was added to sample C. 
There, the  $\delta$-doping between first and second QW was placed asymmetrically 
(see Fig.~\ref{fig1}c) thus introducing a small asymmetry to the whole structure, 
consisting of 10 QWs in total. 
Samples grown along $z\parallel[110]$ were square shaped 
with the sample edges  of 5~mm length oriented along 
$x\parallel[1{\bar 1}0]$ and $y\parallel[00{\bar 1}]$.
To measure photocurrents, ohmic contacts 
were alloyed in the center of each  sample side.

\begin{figure}[t]
\centerline{\epsfxsize 70mm \epsfbox{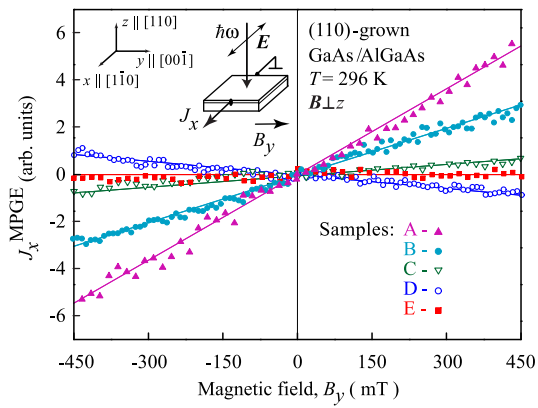}}
\caption{Magnetic field dependences of $J^{{\rm \: MPGE}}_x$ 
for the radiation 
polarized
along $x$ and 
an in-plane magnetic field ($\bm{B} \parallel y$).}
\label{fig2}
\end{figure}

Photocurrents are measured  at room temperature 
by exciting the samples with
linearly polarized  terahertz radiation under normal incidence.
The pulsed THz radiation is obtained  
from an optically pumped pulsed NH$_3$ molecular laser~\cite{Ganichev2006}. 
The wavelength of $148\,\mu$m 
was chosen to cause only $intra$-subband (Drude-like) absorption of
the radiation. The geometry of the experiment is sketched in the
inset of Fig.~\ref{fig2}. The photocurrent is measured in
unbiased structures via the voltage drop across a 50~$\Omega$ load
resistor.
The width of the photocurrent pulses is about 100~ns which corresponds to the 
terahertz laser pulses duration.
In the experiments described below we   
probe the sample asymmetry by means of the  magnetic field
dependence of the photocurrent with the  
polarization vector of the incoming light aligned along the 
$x$-axis. An external magnetic field with a maximum  strength up to $B = 0.5$~T
is applied either in-plane, parallel to $y$, or
normal to QW plane. 
To relate the photocurrent and hence the asymmetry with spin lifetimes $\tau_s$  
of the same devices, time-resolved Kerr rotation (TRKR)
measurements have been performed. 
Here, the samples
are excited by a circularly-polarized laser pulse from a mode-locked
Ti:sapphire laser. The laser energy is tuned to excite
spin-polarized electron-hole pairs where the electrons are slightly above 
the Fermi energy of the 2D
electron system. The spin polarization induced in the samples by the
pump pulse is probed by a second, time-delayed linearly-polarized pulse of the same
laser. Due to the magneto-optic Kerr
effect, the polarization axis of the reflected probe beam is changed
by a small angle, proportional to the $z$-component of the spin
polarization in the sample. This Kerr rotation angle is measured by
a balanced detector. Standard lock-in technique is used, in which
the pump beam is modulated by a flywheel chopper. These measurements
are performed at $T$~=~40~K.

\begin{figure}[t]
\centerline{\epsfxsize 70mm \epsfbox{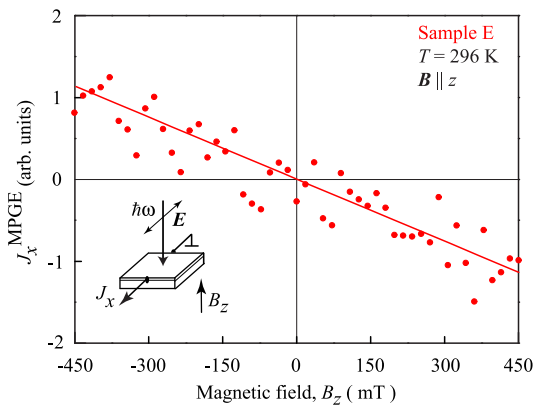}}
\caption{$J^{{\rm \: MPGE}}_x (B)$ 
for sample E measured  
for the radiation 
polarized
along $x$ and a magnetic field
normal to the QWs.
}
\label{fig3}
\end{figure}

The degree of  SIA is reflected in the magnetic field 
dependence of the photocurrent 
displayed in Fig.~\ref{fig2}. To eliminate a magnetic field independent  
photocurrent-offset~\cite{footnote},  
several times smaller than the photosignal at $B_y = \pm$~0.5~T, 
we subtract this background and plot the data in the
form $J^{{\rm \: MPGE}}_x = J_x(B) - J_x(0)$.
The currents shown in Fig.~\ref{fig2}
are directly proportional to the applied field but 
the slope of $J^{{\rm \: MPGE}}_x(B_y)$ is sample dependent.
The largest  slope is obtained for
sample A with the strongest asymmetry while the slope vanishes for the symmetric sample E. 
In case of sample D, having negative $\chi = (l-r) / (l+r)$, also the slope is negative. 
The situation changes in the presence of an out-of-plane 
magnetic field $B_z$. For this configuration the MPGE current 
is observed for all structures including sample E. The result for sample E is plotted
in Fig.~\ref{fig3}.

Before continuing we show within the framework of the phenomenological theory~\cite{Belkov05}
how symmetry is connected with the MPGE effect. 
As the effect is caused by the lack of an inversion symmetry, 
both, SIA and BIA  contribute to the observed phenomena. 
It turns out that in case of an in-plane external magnetic field the MPGE signal 
develops only in asymmetric QWs while it is absent
in symmetrical QWs having the higher point group symmetry  $C_{2v}$.
This effect allows us to analyze the symmetry of the system. 
Phenomenological theory of the MPGE
gives for linearly polarized radiation at normal incidence~\cite{Belkov05}
\begin{equation} \label{phen0}
j^{{\rm \: MPGE}}_\alpha = \sum_{\beta\gamma\delta}
\phi_{\alpha\beta\gamma\delta}\:B_\beta\:\frac{\left(E_\gamma
E^*_\delta + E_\delta  E^*_\gamma\right)}{2}\:.
\end{equation}
Here $\bm{\phi}$ is a fourth rank pseudo-tensor being symmetric in
$\gamma$ and $\delta$, $B_{\beta}$ and $E_\gamma$ are the components of
the magnetic field $\bm{B}$ and 
radiation electric field $\bm E$, respectively.

MPGE requires  linear coupling of an axial vector, magnetic field, 
and an in-plane polar vector, current  (see Eq.~(\ref{phen0})), 
and occurs for certain $\bm{B}$-field orientations when for all symmetry 
operation the vectors transforms are the same. 
While the symmetry of 
perfectly symmetric (110)-grown QWs belongs 
to the point group  $C_{2v}$, 
asymmetric QWs belongs to the point group $C_s$. 
The point group $C_s$ contains only two symmetry elements:
identity and a mirror plane $m_1$, perpendicular to the $x$-axis. 
For $C_s$ group
symmetry requirements are fulfilled for $j_x$ and $B_y$ or $B_z$ only. 
Indeed reflection by the  $m_1$-plane reverses the sign of 
$j_x$ ($j_x \rightarrow -j_x$) and two components of the magnetic field $\bm B$
($B_y \rightarrow -B_y$ and $B_z \rightarrow -B_z$). 
Thus the MPGE can occur for  magnetic fields aligned in-plane 
and out-of-plane of the QW. The photocurrent for $\bm{E} \parallel x$, as used in experiments,
is given by
\begin{equation}\label{MPGE_Cs}
j_{x}^{{\rm \: MPGE,}\:C_{s}} 
= \phi^{SIA}_{xyxx}\:B_y\,|E_x|^2 + \phi^{BIA}_{xzxx}\:B_z\,|E_x|^2\:
\end{equation}
with tensors components $\phi^{SIA}_{xyxx}$ and $\phi^{BIA}_{xzxx}$
determined by the degree of the SIA and BIA, respectively. If  
the magnetic field is applied in $y$-direction 
the last term in Eq.~(\ref{MPGE_Cs}) becomes zero 
and the MPGE current is determined solely by the SIA 
coefficient. The latter determines the degree of SIA and is closely 
related to our 
parameter  $\chi$.

The experiment displayed in Fig.~\ref{fig2} 
(samples A to E) shows that the magnitude of the $J(B_y)$-slope strongly 
depends on the  doping profile. Furthermore,
if the sign of $\chi$ is reversed (from sample B to D), the slope of the 
photocurrent gets reversed, too (see 
Fig.~\ref{fig2}).  
As the 
MPGE 
current is proportional to the SIA coefficient, 
these observations demonstrate that the position of the doping layer
can be effectively used for tuning  the structure  
asymmetry strength. In particular,
the sign of  $\phi^{SIA}_{xyxx}$ can be inverted by putting the doping layer, 
keeping the distance fixed, to the other side of the QW. 
From this simple $\phi^{SIA}_{xyxx}$ 
to 
$\chi$   relation one would expect that the magnitudes 
of the  $J(B_y)$-slopes of samples B and D are the same. 
However the magnitudes of two slopes differ by a factor of 2.
This can be attributed to subtle details of impurity incorporation
during growth. 
We saw that samples 
B and D
grown in the similar way 
but with the reversed doping positions 
differ by a factor 4 in electron mobility.
We ascribe the smaller slope 
observed in sample D
to increased scattering which reduces the photocurrent.
Equation~(\ref{MPGE_Cs}) gives also the MPGE current for a magnetic field applied 
normal to the QW plane, as observed in experiment.

While for an in-plane magnetic field the photogalvanic effect, 
described by Eq.~(\ref{phen0}), is observable in asymmetrical structures, it is forbidden in symmetrically 
grown QWs with the higher point group symmetry  $C_{2v}$.
This is caused by an additional mirror plane $m_2$ being parallel to the 
QW plane of symmetrically grown (110)~structures. 
Indeed, reflection by this  plane does not modify $j_x$ 
but changes the polarity of in-plane axial vector $\bm B$. 
Therefore in such systems linear coupling of $j_x$  and $B_y$ is forbidden.
On the other hand, mirror reflection of plane $m_2$ does not 
modify the $z$-component of $\bm B$. Thus the coupling of $j_x$ and 
$B_z$ is allowed for reflections by both, $m_1$ and $m_2$ planes,
and photocurrent $j_{x}$ can occur in symmetric (110)-oriented QWs
in the presence of a magnetic field in $z$-direction. 
For this symmetry Eq.~(\ref{phen0}) is reduce to
\begin{equation}\label{MPGE_C2v}
j_{x}^{{\rm \: MPGE,}\:C_{2v}} =  \phi^{BIA}_{xzxx}\:B_z\,|E_x|^2\:.
\end{equation}
This equation
fully describes  
the data taken from sample E
(see Figs.~\ref{fig1} and \ref{fig2}). 
There, no MPGE is observed for 
in-plane magnetic field
but a sizeable 
effect is detected for 
$\bm{B}$
applied normal to the QW plane. 
In case of sample E, the absence of a magnetic field induced photocurrent 
in an in-plane $\bm{B}$  indicates that the QW is highly symmetric and 
lacks the structure  asymmetry.
The signal, observed in the same structure for an out-of-plane $B_z$-field 
stems from the BIA term (see Eq.~(\ref{MPGE_C2v})). 
Hence measurement of the MPGE gives us an experimental handle
to analyze the degree of SIA.

\begin{figure}[t]
\centerline{\epsfxsize 70mm \epsfbox{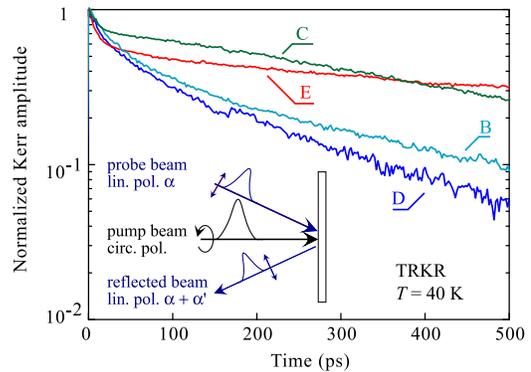}}
\caption{Kerr rotation measured in  
asymmetrically grown QWs samples B, C, D and symmetrical structure E.} 
\label{fig4}
\end{figure}

The structure inversion asymmetry determines  the Rashba spin splitting 
and therefore controls
the D'yakonov-Perel' spin relaxation~\cite{DP71} along the $z$-direction. Any
variation of SIA, e.g., due to asymmetric doping,  should result in a variation of 
the spin relaxation time. 
To directly demonstrate this connection, we compare spin relaxation rates measured 
in the symmetrically doped QW sample E and the 
asymmetrically doped QW samples B, C and D. 
We extract the spin lifetime $\tau_s$  from time-resolved Kerr rotation measurements.
The time evolution of the Kerr rotation angle tracks the spin polarization
within the sample. By fitting an exponential decay function to the
data, $\tau_{s}$ is determined. In correspondence to the above  photocurrent measurements
indicating a larger degree of asymmetry of the samples B and D
compared to C and E, we observe that $\tau_{s}$ in sample E
is more than three times larger  than that in sample B and about two
times larger than in sample D (see Table~\ref{table1}). 
We note that in  sample C the spin relaxation time is also rather long. In 
this structure  the $\delta$-doping for 9 QWs  is placed in the
center of each barrier separating the QWs and only the first QW is asymmetric. While the 
photocurrent measurements reflect the asymmetry of the whole structure by measuring
the Kerr rotation  we probe spin relaxation only of selected QWs (in our case 9 symmetrical QWs):
Due to the change of the potential profile the asymmetric QW has 
an effective bandgap, which is different from the symmetric QWs and thus does not 
contribute significantly to the Kerr rotation signal detected at 
interband resonance conditions with the symmetric QWs.

The relaxation of  spins oriented along $z$-axis due to the D'yakonov-Perel'
mechanism in (110)-grown QWs is caused \textit{solely} 
by SIA~\cite{DK86}. The measurements on symmetrical sample E and almost symmetrical sample C 
reproduce the results obtained by Ohno et~al.~\cite{Ohno1999}, where long spin lifetimes were found for
nominally undoped (110)-grown QWs by time- and polarization-resolved
transmission measurements, as well as those obtained by D\"ohrmann
et~al.~\cite{x3}  by means of  time- and polarization-resolved
photoluminescence (TRPL) on symmetrically grown \textit{n}-doped (110)~QWs.
We note that in  both the TRKR and the TRPL  measurements,
electron-hole pairs are generated in order to probe $\tau_{s}$.
Thus, the Bir-Aronov-Pikus mechanism of spin relaxation, in which
the randomly fluctuating hole spins couple to the electron spins~\cite{BAP76}, 
may not be fully neglected. 
The difference in $\tau_{s}$ between structures B and D having large asymmetry 
stems from both the different electron mobility and the degree of asymmetry of these samples:
In the motional-narrowing regime of D'yakonov-Perel' spin
relaxation, $\tau_{s}$ is inversely proportional to the momentum
relaxation time, which is correlated with mobility.

As an important result of all our measurements we 
obtained the zero current response and the
longest spin relaxation time from the almost symmetrically
doped QWs
which set an upper limit of spin dephasing in GaAs QWs.
This is in contrast to (001)-grown
structures where such a 
geometry results in a
substantial SIA~\cite{prBRD}. This essential difference
stems from the growth temperature, and,
subsequently, the diffusion length.
Indeed, the growth temperature of high-quality
(001)-oriented QWs is higher than 600$^\circ$C,
while (110)-structures are grown at 
480$^\circ$C~\cite{W2}. High growth
temperature of (001)-oriented 
heterostructures 
leads to substantial dopant
migration and results in structure asymmetry of symmetrically doped
QWs. The investigation of 
MPGE,
in particular the sign 
inversion by the reversing of structural
asymmetry and the zero response of 
symmetrical 
structures,
provides an effective access to  study  the symmetry of 
(110)-oriented QWs.
Summarizing, our measurements 
of the photocurrent give the
necessary feedback to 
reliably growth structures with long spin relaxation times.

We thank E.~L.~Ivchenko
for helpful discussions.  This work is supported by the
DFG via programms
SPP~1285, SFB~689 and GRK~638,
Russian President grant for young scientists, RFBR,  
and RSSF.


\begin{widetext}
\begin{table*}
\caption{\label{table1}Parameters of samples and spin lifetimes for the
symmetrically- and asymmetrically-grown QWs. 
}
\begin{ruledtabular}
\begin{tabular}{ccccccccc}
\hline
 sample &   \,\,\,\, spacer 1 & \,\,\,\,  QW width  & \,\,\,\, spacer 2 &\,\,\,\, $\chi = \frac{l-r}{l+r}$ & \,\,\,\, mobility at 4.2~K & \,\,\,\, density at 4.2~K
&  \,\,\,\, $\Delta J^{MPGE}_x/ \Delta B_y $ &  \,\, $\tau_{s}$ at 40~K  \\

  &              \,\,   $l$ (nm)  &   \,\, $L_W$ (nm)        &  \,\,  $r$ (nm) &   &  \,\,(cm$^2$/Vs) &  (cm$^{-2}$) & (arb. units)  &    \,\,  (ps) \\
\hline

A 
  &  $-$    &  $\,\,-$     & 50.0 & $-$ & 1.4 $\times 10^5$ &  1.4 $\times 10^{11}$ &\, 18.6 & $-$  \\

B
  & 165.0      &  15.0    & 20.0  &\, 0.78 & 7.9 $\times 10^4$  & 2.6  $\times 10^{11}$ &\, 6.6 & 326 $\pm$ 10   \\

\,\,C 
  & 92.5      &    15.0 &  92.5 & \,\,0.06 & 1.8 $\times 10^5$ &  0.9 $\times 10^{11}$  &\, 1.5 & 500  $\pm$ 15   \\

D
  & 20.0      &    15.0   & 165.0   & -0.78 & 3.4 $\times 10^5$ &  3.4 $\times 10^{11}$ & -2.7 &192 $\pm$ 9  \\

E
  &  40.0     & 20.0     &   40.0 & \,\,0 &  $-$ &   $-$ & \,0 & 638 $\pm$ 2   \\
\hline
\end{tabular}
\end{ruledtabular}
\end{table*}
\end{widetext}


\begin{thebibliography}{99}
\bibitem{footnote} The background signal is caused by linear photogalvanic 
effect~\cite{Ivchenkobook2}, which is out of scope of this letter.
\end{thebibliography}
\end{document}